\documentclass[preprint]{elsarticle}
\usepackage{amsmath,amssymb,graphicx, amsthm}
\newtheorem{Thm}{Theorem}

\newtheorem{Lem}{Lemma}
\newtheorem{Rmk}{Remark}

\begin{document}

\begin{frontmatter}
\journal{SISC}

\title{ Discrete Conservation Law on Curved Surfaces}

 \author[Chen]{Sheng-Gwo Chen\corref{corresponding}}
 \author[Wu]{Jyh-Yang Wu}\ead{jywu@math.ccu.edu.tw}

\cortext[corresponding]{Corresponding author email :
csg@mail.ncyu.edu.tw}
\address[Chen]{Department of Applied Mathematics, National Chiayi University,  Chia-Yi 600,
Taiwan.}
\address[Wu]{Department of Mathematics, National Chung Cheng University, Chia-Yi 621,
Taiwan.}

\begin{abstract}
In this paper we shall introduce a simple, effective numerical
method for finding differential operators for scalar and
vector-valued functions on surfaces. The key idea of our algorithm
is to develop an intrinsic and unified way to compute directly the
partial derivatives of functions defined on triangular meshes which
are the discretization of regular surfaces under consideration. Most
importantly, the divergence theorem and  conservation laws on
triangular meshes are fulfilled.
\end{abstract}

\end{frontmatter}

\section{Introduction}
Numerical methods to compute partial differential operators on
regular surfaces have always received great interest over last
decades. However, they are still not well-understood. For example,
Conservation laws for diffusion equations are usually unsatisfied.
Conservation Law is an important principle in physics. Indeed,
Conservational laws plays a key role in the study of partial
differential equations and have many applications in the
linearization, integrability and numerical analysis. The solution
$u(x,t)$ of diffusion equation
\begin{equation}
u_t = \alpha \Delta_{\Sigma} u
\end{equation}
with $\alpha > 0 $  on a regular surface $\Sigma$ preserves the
total energy. That is,
\begin{equation}
\frac{\partial}{\partial t} \int_{\Sigma} u(x,t) dx = 0.
\end{equation}

To numerically simulate $u(x,t)$, one discretizes the regular
surface $\Sigma$ to obtain a triangular surface mesh $S$ of
$\Sigma$, and considers the discrete solution $u(x,t)$ on $S$. In
this way, a fundamental problem arise. Usually, the solution
$u(x,t)$ on $S$ will not preserve the total energy. That is,
\begin{equation}
\int_{\Sigma} u(x,t_i)
\end{equation}
will change as $t_i$ increases.

The violation of the Conservation Law comes from the discretization
of the Laplacian-Beltrami operator $\Delta_{\Sigma}$ on $\Sigma$. In
this paper, we shall try to handle this defect. Lai et. al.
\cite{Lai} discussed this problem for regular curves in 2008.

Partial differential equations (PDEs) need to be solved
intrinsically and numerically for data defined on 3D regular
surfaces in many applications. For instance, such examples exist in
fluid dynamic flows (Diewald, Preufer and Rumpf \cite{Diewald}),
(Bertalmio, Cheng, Osher and Sapiro \cite{Bertalmio}),texture
synthesis (Turk\cite{Turk}, Witkin and Kass\cite{Witkin}), vector
field visualization (Diewald, Preufer and Rumpf\cite{Diewald}),
weathering (Dorsey and Hanrahan\cite{Dorsey}) and cell-biology
(Ayton, McWhirter, McMurty and Voth\cite{Ayton}). Usually, regular
surfaces are presented by triangular or polygonal forms. Partial
differential equations are then solved on these triangular or
polygonal meshes with data defined on them. The use of triangular or
polygonal meshes is very popular in all areas dealing with 3D
models. However, it has not yet been a widely accepted method to
compute differential characteristics such as principal directions,
curvatures and Laplacians (Chen and Wu\cite{Chen1,Chen2}, Wu, Chen
and Chi\cite{Wu}, Taubin\cite{Taubin}). In Wu, Chen and
Chi\cite{Wu}, the authors proposed a new intrinsic simple algorithm,
LTL method, to handle this difficulty. In this note, we shall use
this new technique to approximate gradient, divergence and
Laplace-Beltrami operators on surfaces.

In Osher and Sethian\cite{Osher} and Bertalmio, Cheng, Osher and
Sapiro\cite{Bertalmio} discussed a framework, the implicit surface
algorithm, to solve Variational problems and PDE's for scalar and
vector-valued data defined on regular surfaces. Their key idea is to
use, instead of a triangular or polygonal representation, an
implicit representation. The surface under consideration is the
zero-level set of a higher dimensional embedding function. Then they
smoothly extend the original data on the surface to the 3D domain,
adapt the PDE's accordingly, and implement all the numerical
computations on the fixed Cartesian grid corresponding to the
embedding function. The advantage of their method is the use of the
Cartesian grid instead of a triangular mesh for the numerical
implementation.

The discretizations of the gradient, divergence and Laplace-Beltrami
operators that we will discussin this paper will have the following
advantages:
\begin{description}
 \item[Intrinsicness:] we use the intrinsic geometric LTL method to define these operators.
 \item[Conservation:] our Laplace-Beltrami operator will satisfy   conservation laws on triangular meshes for diffusion equations.
 \item[Convergence:] our gradient, divergence and Laplace-Beltrami operators will have the linear convergence rate locally and uniformly.
 \item[Simplicity:] the numerical computations are also very easy to implement.
\end{description}

The rest of this paper is organized as follows. In section 2, we
recall the gradient, divergence and Laplace-Beltrami operators
defined on regular surfaces. In section three we propose our new
discrete algorithm for these differential operators on triangular
meshes. We also discuss the convergence problem and conservation
laws for these operators. Numerical simulations are presented in
section 4.

\section{The gradient, divergence and LB operators on regular surfaces}

In order to describe the gradient, divergence and the LB operator on
functions or vector fields in a regular surface $\Sigma$ in the 3D
Euclidean space $\mathbb{R}^3$, we consider a parameterization $ x:
U \rightarrow \Sigma$ at a point $p$, where $U$ is an open subset of
the 2D Euclidean space $\mathbb{R}^2$. We can choose, at each point
$q$ of $x(U)$, a unit normal vector $N(q)$. The map $N : x(U)
\rightarrow S^2$ is the local Gauss map from an open subset of the
regular surface $\Sigma$ to the unit sphere  $S^2$ in the 3D
Euclidean space $\mathbb{R}^3$. The Gauss map $N$ is differentiable.
Denote the tangent space of $\Sigma$ at the point $p$ by $T\Sigma_p
= \{ v \in \mathbb{R}^3 | v \perp N(p) \}$. The tangent space
$T\Sigma_p$  is a linear space spanned by $\{ x_u, x_v\}$  where
$u,v$  are coordinates for $U$. The gradient $\nabla_{\Sigma} g$  of
a smooth function $g$ on $\Sigma$ can be computed from
\begin{equation}\label{gradient_g}
\nabla_{\Sigma} g = \frac{g_uG - g_vF}{EG-F^2} x_u + \frac{g_vE - g_uF}{EG-F^2}x_v
\end{equation}
where $E, F$, and $G$ are the coefficients of the first fundamental form and
\begin{equation}
g_u = \frac{\partial g(x(u,v))}{\partial u} \mbox{ and } g_v = \frac{\partial g(x(u,v))}{\partial v}.
\end{equation}
See do Carmo\cite{Docarmo1}  for the details.

Let  $X = Ax_u + Bx_v$ be a local vector field on $\Sigma$. The
divergence, $\nabla_{\Sigma} \cdot X$, of $X$  is defined as a
function $\nabla_{\Sigma} \cdot X : \Sigma \rightarrow \mathbb{R}$
given by the trace of the linear mapping $Y(p) \rightarrow
\nabla_{Y(p)}X$ for $p \in \Sigma$. A direct computation gives
\begin{equation}\label{divergence_X}
\nabla_{\Sigma}\cdot X = \frac{1}{\sqrt{EG-F^2}}\left(
\frac{\partial }{\partial u } (A\sqrt{EG-F^2} ) + \frac{\partial
}{\partial v } (B\sqrt{EG-F^2} ) \right)
\end{equation}

The LB operator $\Delta _{\Sigma}$ acting on the function $g$ is defined by the integral duality
\begin{equation}
(\Delta_{\Sigma}g, \varphi) = -(\nabla_{\Sigma}g, \nabla_{\Sigma}\varphi)
\end{equation}
for all smooth function $\varphi$ on $\Sigma$. That is,
$\Delta_{\Sigma} g = \nabla_{\Sigma} \cdot \nabla_{\Sigma}g$. A
direct computation yields the following local representation for the
LB operator $\Delta_{\Sigma} g$ on a smooth function $g$:
\begin{equation}\label{laplace_g}
\begin{array}{ll}
\Delta_{\Sigma} g & = \frac{1}{\sqrt{EG-F^2}}\left [ \frac{\partial}{\partial u} (\frac{G}{\sqrt{EG-F^2}} \frac{\partial g}{\partial u} )
-  \frac{\partial}{\partial u} (\frac{G}{\sqrt{EG-F^2}} \frac{\partial g}{\partial v} ) \right ]\cr\cr
& +  \frac{1}{\sqrt{EG-F^2}}\left [ \frac{\partial}{\partial v} (\frac{G}{\sqrt{EG-F^2}} \frac{\partial g}{\partial v} )
-  \frac{\partial}{\partial v} (\frac{G}{\sqrt{EG-F^2}} \frac{\partial g}{\partial u} )  \right ]
\end{array}
\end{equation}

\section{Discrete gradient, divergence and LB operators}

In this section we shall describe a simple and effective method to
define the discrete gradient, divergence and LB operator on
functions or vector fields on a triangular mesh. The primary ideas
were developed in Chen, Chi and Wu\cite{Chen1,Wu} where we try to
estimate the discrete partial derivatives of functions on 2D
scattered data points. Indeed, the method that we shall use to
develop our algorithm is divided into two main steps: first we lift
the 1-neighborhood points to the tangent space and obtain a local
tangential polygon. Second, we use some geometric idea to lift
functions or vectors to the tangent space. We call this a local
tangential lifting (LTL) method. Then we present a new algorithm to
compute their gradients in the 2D tangent space. This means that the
LTL process allows use to reduce the 2D curved surface problem to
the 2D Euclidean problem.

Consider a triangular surface mesh $S=(V,F)$, where $V=\{v_i | 1\leq
i \leq n_V\|$ is the list vertices and  $F=\{T_k | 1\leq k \leq
n_F\}$ is the list of triangles.

\subsection{The local tangential lifting (LTL) method}
To describe the local tangential lifting (LTL) method, we introduce
the local tangential polygon   at a vertex  $v$ of $V$ as follows:
\begin{enumerate}
\item The normal vector $N_A(v)$ at the vertex $v$  in $S$  is given by
\begin{equation}\label{weighted_normal}
N_A(v) = \frac{\sum\limits_{T \in T(v)} \omega_T N_T}{\|\sum\limits_{T \in T(v)} \omega_T N_T\|}
\end{equation}

where $N_T$  is the unit normal to a triangle face $T$  and the centroid weight is given in [3] by
\begin{equation}\label{centroid_weight}
\omega_T = \frac{\frac{1}{\|G_T - v\|^2}}{\sum\limits_{\tilde{T} \in T(v)} \frac{1}{\|G_{\tilde{T}} - v\|}}
\end{equation}
Here,  $G_T$ is the centroid of the triangle face $T$  determined by
\begin{equation}
G_T = \frac{v+v_i+v_j}{3}.
\end{equation}
Note that the letter A in the notation  $N_A(v)$ stands for the word "Approximation".

\item   The approximating tangent plane $TS_A(v)$ of $S$  at $v$   is now determined by
$TS_A(v) =\{ w \in \mathbb{R}^3 | w \perp N_A(v)\}$.

\item The local tangential polygon $P_A(v)$ of $v$  in $TS_A(v)$   is formed by the vertices $\bar{v}_i$   which is the lifting vertex  of $v_i$   adjacent to $v$  in $V$.
\begin{equation}
\bar{v}_i = (v_i - v) - <v_i - v, N_A(v)>N_A(v)
\end{equation}
as in figure \ref{tangential_polygon}.

\begin{figure}
{\center
\includegraphics[width= .5\textwidth]{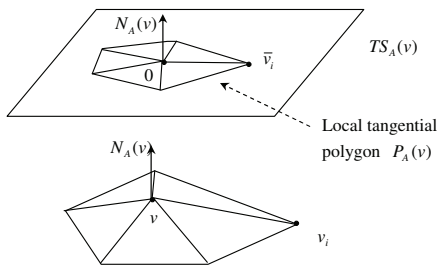}
\caption{The local tangential polygon $P_A(v)$ }\label{tangential_polygon}
}
\end{figure}

\item We can choose an orthonormal basis $e_1, e_2$  for the tangent
plane  $TS_A(v)$  of $S$  at $v$  and obtain an orthonormal
coordinates $(x,y)$ for vectors  $w \in TS_A(v)$ by $w =
xe_1+ye_2$. We set $\bar{v}_i = x_ie_1+y_ie_2$  with respect to
the orthonormal basis $e_1, e_2$.

\end{enumerate}

Next we explain how to lift locally a function defined on $V$  to
the local tangential polygon $P_A(v)$. Consider a function $h$ on
$V$. We will lift locally the function $h$  to a function of two
variables, denoted by $\bar{h}$, on the vertices $\bar{v}_i$  in
$P_A(v)$ by simply setting
\begin{equation}
\bar{h}(x_i, y_i) = h(v_i).
\end{equation}
and $\bar{h}(\vec{0})$  where $\vec{0}$  is the origin of $TS_A(v)$.
Then one can extend the function $\bar{h}$  to a piecewise linear
function on the whole polygon $P_A(v)$   in a natural and obvious
way.

\subsection{A new discrete gradient algorithm}
In this subsection we present a new discrete 2D algorithm for the
gradients of functions on the 2D domains in the $x-y$ plane and also
on triangular surface meshes. Given a $C^3$  function $f$  on a
domain $\Omega$  in the $x-y$ plane with the origin $(0,0) \in
\Omega$, Taylor's expansion for two variables $x$ and $y$ gives
\begin{equation}
\begin{array}{ll}
f(x,y) & = f(0,0) + xf_x(0,0) + yf_y(0,0) \cr
    & + \frac{x^2}{2} f_{xx}(0,0) + xy f_{xy}(0,0) +\frac{y^2}{2}f_{yy}(0,0) + O(r^3)
\end{array}
\end{equation}
when $r=\sqrt{x^2+y^2}$  is small.

Consider a family of neighboring points $(x_i, y_i) \in \Omega$,
$i=1,2,\cdots,n$, of the origin $(0,0)$. Take some constants
$\alpha_i$, $i=1,2,\cdots, n$ with $\sum_{i=1}^n \alpha^2_i = 1$.
Then one has
\begin{equation}
\begin{array}{l}
\sum_{i=1}^n\alpha_i(f(x_i,y_i) - f(0,0)) \cr
      = (\sum_{i=1}^n \alpha_i x_i) f_x(0,0) + (\sum_{i=1}^n \alpha_i y_i) f_y(0,0)
       + \frac{1}{2}(\sum_{i=1}^n \alpha_i x_i^2) f_{xx}(0,0) \cr
       +   (\sum_{i=1}^n \alpha_i x_i y_i) f_{xy}(0,0)   + \frac{1}{2}(\sum_{i=1}^n \alpha_i y_i^2) f_{yy}(0,0) + O(r^3).
\end{array}
\end{equation}

We choose the constants $\alpha_i$, $i=1,2,\cdots, n$  so that they satisfy the following equations:
\begin{description}
\item[(i)] $\sum_{i=1}^n \alpha_i x_i y_i = 0$
\item[(ii)] $\sum_{i=1}^n \alpha_i x_i^2 = 0$
\item[(iii)] $\sum_{i=1}^n \alpha_i y_i^2 = 0$
\end{description}
One can rewrite these equations in a matrix form and obtain
\begin{equation}\label{system_gradient}
\begin{pmatrix}
x_1y_1, & x_2y_2, & \cdots \cdots, & x_ny_n \cr
x_1^2, & x_2^2, & \cdots \cdots, &x_n^2 \cr
y_1^2, & y_2^2, & \cdots \cdots, &y_n^2
\end{pmatrix}
\begin{pmatrix}
\alpha_1 \cr \alpha_2 \cr \vdots \cr \alpha_n \end{pmatrix} = \begin{pmatrix} 0 \cr 0 \cr 0\end{pmatrix}
\end{equation}
Therefore, we have, for these solutions  $\alpha_i$,
\begin{equation}
\sum_{i=1}^n \alpha_i (f(x_i,y_i) - f(0,0)) = (\sum_{i=1}^n \alpha_i x_i) f_x(0,0) + (\sum_{i=1}^n \alpha_i y_i) f_y(0,0) + O(r^3)
\end{equation}
Choose another solutions $\beta_i$, $i=1,2,\cdots,n$, with $\sum_{i=1}^n \beta_i^2 = 1$  for the linear system (\ref{system_gradient}). We have
\begin{equation}
\begin{array}{l}
(\sum_{i=1}^n \alpha_i x_i) f_x(0,0) + (\sum_{i=1}^n \alpha_i y_i) f_y(0,0) = \sum_{i=1}^n \alpha_i (f(x_i,y_i) - f(0,0)) + O(r^3) \cr
(\sum_{i=1}^n \beta_i x_i) f_x(0,0) + (\sum_{i=1}^n \beta_i y_i) f_y(0,0) = \sum_{i=1}^n \beta_i (f(x_i,y_i) - f(0,0))+ O(r^3)
\end{array}
\end{equation}
If the valence $n$ of the origin $(0,0)$ is at least $5$, we can
choose the solutions $\alpha_i$  and $\beta_i$ so that the following
coefficient matrix is invertible.
$$ \begin{pmatrix}
\sum_{i=1}^n \alpha_ix_i, & \sum_{i=1}^n \alpha_iy_i \cr
\sum_{i=1}^n \beta_ix_i, & \sum_{i=1}^n \beta_iy_i
\end{pmatrix} $$
Under these circumstances, we can find the gradient $\nabla f(0,0) = (f_x(0,0), f_y(0,0))$
by the relation
\begin{equation}\label{gradient_f}
\begin{pmatrix}
f_x(0,0) \cr
f_y(0,0)
\end{pmatrix} =
\begin{pmatrix}
\sum_{i=1}^n \alpha_i x_i, & \sum_{i=1}^n \alpha_i y_i \cr
\sum_{i=1}^n \beta_i x_i, & \sum_{i=1}^n \beta_i y_i
\end{pmatrix}^{-1}
\begin{pmatrix}
\sum_{i=1}^n \alpha_i (f(x_i,y_i) - f(0,0))  \cr
\sum_{i=1}^n \beta_i (f(x_i,y_i) -f(0,0))
\end{pmatrix}
+O(r^2)
\end{equation}

Next we discuss how to approximate the gradient of a function on
regular surfaces. Let $\Sigma$  be a regular surface and  $S=(V,F)$
a triangular surface mesh of $\Sigma$   with mesh size $r > 0$.
Consider a vertex $v \in V$. The local tangential polygon $P_A(v)$
of $v$  in $T_A(v)$ is formed by the vertices $\bar{v}_i$ which is
the lifting vertex of $v_i$ adjacent to $v$  in $V$. Note that
\begin{equation}
\bar{v}_i=(v_i - v)-<v_i - v,N_A(v)>N_A(v).
\end{equation}
Choose and fix an orthonormal basis $e_1, e_2$ for the tangent plane
$TS_A(v)$ of $S$ at $v$  and obtain an orthonormal coordinates
$(x,y)$ for vectors $w \in TS_A(v)$ by $w = xe_1+ye_2$. We set
$\bar{v}_i = x_ie_1+y_ie_2$ with respect to the orthonormal basis
$e_1, e_2$. Consider a function $h$ on $V$. We will lift locally the
function $h$  to a function of two variables , denoted by $f$, on
the vertices $\bar{v}_i$  in $P_A(v)$  by simply setting
\begin{equation}
f(x_i,y_i) = h(v_i)
\end{equation}
and $f(0,0)=h(v)$  where  $(0,0)$  is the origin of $TS_A(v)$. In this way, we can define the approximating gradient by
\begin{equation}
\nabla_A h(v)=
\begin{pmatrix}
\sum_{i=1}^n \alpha_i x_i, & \sum_{i=1}^n \alpha_i y_i \cr
\sum_{i=1}^n \beta_i x_i, & \sum_{i=1}^n \beta_i y_i
\end{pmatrix}^{-1}
\begin{pmatrix}
\sum_{i=1}^n \alpha_i (f(x_i,y_i) - f(0,0))  \cr
\sum_{i=1}^n \beta_i (f(x_i,y_i) -f(0,0))
\end{pmatrix}
\end{equation}
where $\alpha_i, \beta_i$  can be computed from Equations
(\ref{system_gradient}).

Since the approximating normal vector satisfies $N_{\Sigma} (v) = N_A(v) + O(r^2)$,
one can tell from Equations (\ref{system_gradient})-(\ref{gradient_f}) and obtain easily the following convergence theorem.

\begin{Thm}\label{theorem1} ({\bf Convergence Theorem 1 }) \\
 Under the notations as above, one has
 \begin{equation}
 \nabla_{\Sigma} h(v) = \nabla_A h(v) + O(r^2).
 \end{equation}
\end{Thm}

\subsection{A new discrete divergence algorithm on triangular meshes}
\begin{figure}[htb]
{\center
\includegraphics[width=.8\textwidth]{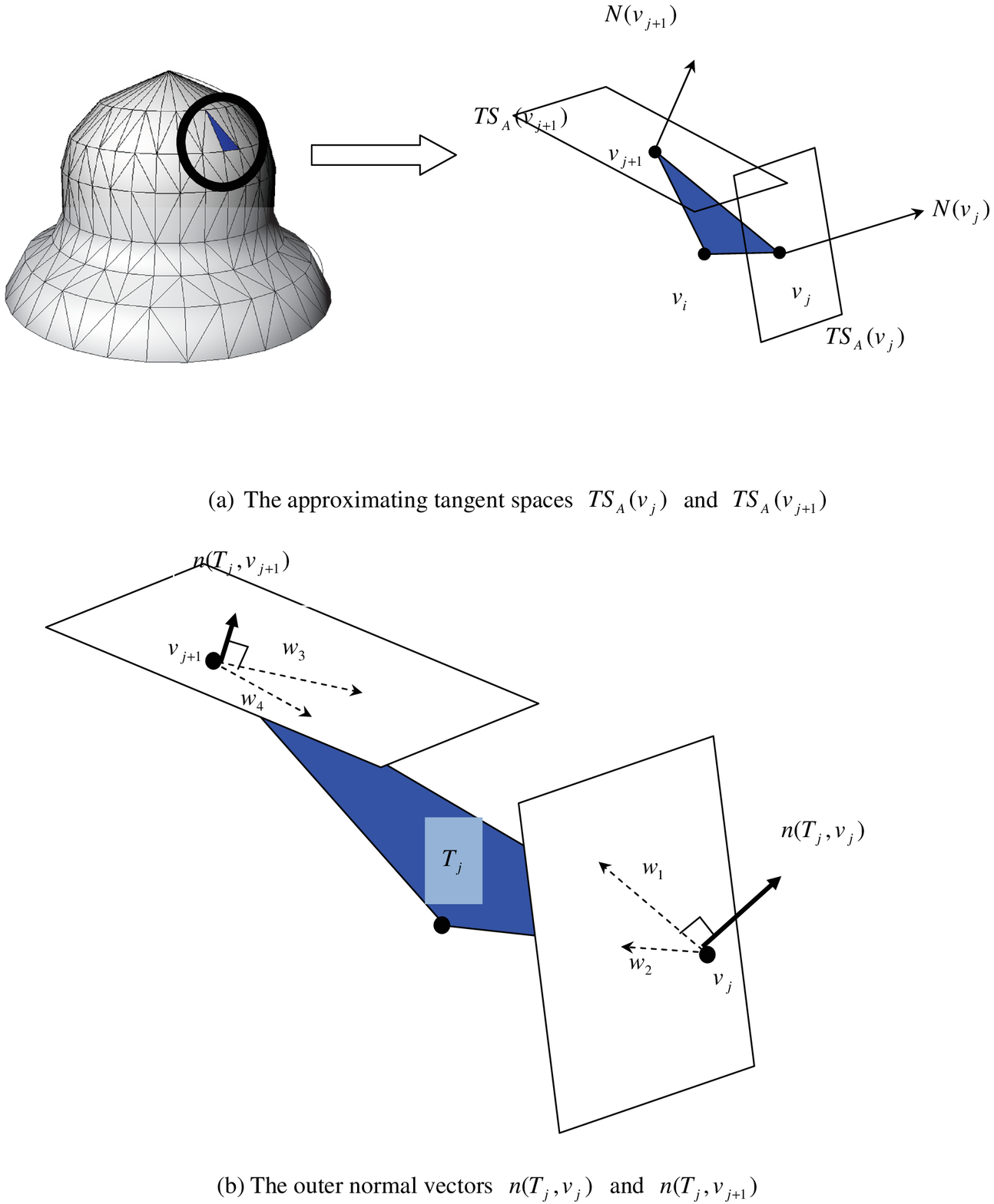}
\caption{The outer normal vectors $n(T_j, v_j)$ and $n(T_j, v_{j+1})$ }\label{outernormals}
}
\end{figure}

In this subsection, we shall use the divergence theorem to give a
discrete approximation of the divergence of a vector field $X$
defined on a triangular surface mesh $S=(V,F)$. Consider a vertex $v
\in V$  and let $v_j$, $j=0,1,\cdots, n$ be the neighboring vertices
of $v$  with $v_0 =v_n$. These vertices $v_j$are labeled
counterclockwise about the normal vector $N_A(v)$.  Let  $T_j$  be
the triangle with vertices $v, v_j$  and $v_{j+1}$. We define the
outer normal vectors  $n(T_j,v_j)$  and $n(T_j,v_{j+1})$ of the
triangle $T_j$  at the vertex $v_j$   and $v_{j+1}$ respectively as
follows. See figure \ref{outernormals}

Consider the lifting vectors $w_1$ and $w_2$ of the vectors $v_{j+1}
- v_j$   and $v - v_j$ to the approximating tangent space $TS_A(v)$
of $S$  at the vertex $v_j$:
\begin{equation}
\left \{
\begin{array}{ll}
w_1 = & (v_{j+1}-v_j) - <v_{j+1} - v_j, N_A(v_j)> N_A(v_j) \cr
w_2 = & (v - v_j) - <v - v_j, N_A(v_j)> N_A(v_j)
\end{array}\right.
\end{equation}
Now the outer normal vector $n(T_j,v_j)$  of the triangle $T_j$  at the vertex $v_j$ can be defined as
\begin{equation}
n(T_j,v_j) = \frac{<w_2,w_1>w_1 - \|w_1\|^2w_2}{\|<w_2,w_1>w_1 - \|w_1\|^2w_2\|}
\end{equation}

Similarly, we consider the lifting vectors $w_3$ and $w_4$  of the
vectors $v_j - v_{j+1}$  and $v - v_{j+1}$  to the approximating
tangent space $TS_A(v_{j+1})$   of $S$  at the vertex $v_{j+1}$:
\begin{equation}
\left \{
\begin{array}{ll}
w_3 = & (v_j-v_{j+1}) - <v_j-v_{j+1}, N_A(v_{j+1})> N_A(v_{j+1}) \cr
w_4 = & (v - v_{j+1}) - <v - v_{j+1}, N_A(v_{j+1})> N_A(v_{j+1})
\end{array}\right.
\end{equation}
Note that the outer normal vectors  $n(T_j,v_{j+1})$ and $n(T_{j+1},v_{j+1})$  are different.

Under these notations, we can now define the discrete divergence $\mathrm{Div}_A X$   of a vector field on the triangular surface mesh $S$ by
\begin{equation}\label{discrete_divergence}
\begin{array}{ll}
\mathrm{Div}_A X(v) = & \frac{1}{\sum_{k=0}^{n-1}|T_k|} \left [  \sum_{j=0}^{n-1} \frac{\|v_{j+1} - v_j \|}{6} ( 2<X(v_j), n(T_j,v_j)> \right. \cr
&  + 2<X(v_{j+1}), n(T_j,v_{j+1})>  + <X(v_j), n(T_j,v_{j+1})> \cr
& \left.  + <X(v_{j+1}), n(T_j,v_j)> ) \right ]
\end{array}
\end{equation}
where $|T_j|$ denotes the area of the triangle $T_j$.

We can extend the divergence $\mathrm{Div}_A X$  to the whole mesh
$S$  piecewise linearly in a natural way. Therefore, we have the
following lemma.
\begin{Lem}\label{lemma1}
Let $X$  be a vector field on the triangular mesh $S$. The integration of the divergence $\mathrm{Div}_A X$ over  $S$  is
\begin{equation}
\int_S \mathrm{Div}_A X = \sum_{v_i \in V} \left [ \frac{1}{3} \mathrm{Div}_A X(v_i) \sum_{j \in N(i) }|T_j| \right ].
\end{equation}
\end{Lem}

Lemma \ref{lemma1} along with the definition of $\mathrm{Div}_A X$  gives the following result.

\begin{Thm}\label{theorem2} ({\bf Discrete Conservation Law 1} ) \\
 Let  $S=(V,F)$ be a triangular mesh without boundary and $X$  a vector field on $S$. We have
 \begin{equation}
 \int_S \mathrm{Div}_A X = 0.
 \end{equation}
\end{Thm}

\begin{figure}
{\center
\includegraphics[width=.8\textwidth]{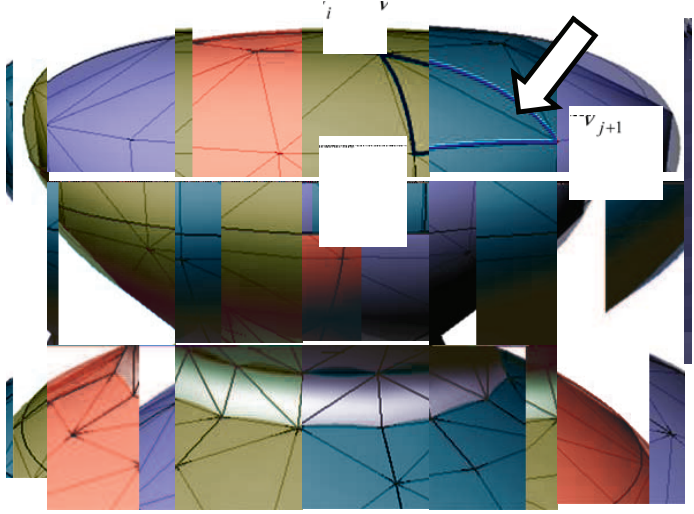}
\caption{ The geodesic triangle $\tilde{T}$ (the blue surface bounded by red curves) n the surface $\Sigma$ (blue surface) }\label{geodesic_triangle}
}\end{figure}

Let  $\Sigma$ be a regular surface and  $S=(V,F)$ a triangular
surface mesh of $\Sigma$  with mesh size $r>0$. If the mesh size $r$
is sufficiently small, we can find a unique geodesic $\gamma_{ij}$
joining two adjacent vertices $v_i$  and $v_j$. In this way, every
triangle $T \in F$  has a corresponding geodesic triangle
$\tilde{T}$  on the surface $\Sigma$ with the same vertices as  $T$.
See figure \ref{geodesic_triangle}. Since the approximating normal
vector satisfies $N_{\Sigma}(v) = N_A(v)+O(r^2)$, the outer normal
vectors $n(\tilde{T}_j,v_j)$  and $n(\tilde{T}_j,v_{j+1})$  of the
geodesic triangle $\tilde{T}_j$  in $\Sigma$  at the vertices $v_j$
and $v_{j+1}$   respectively also have the relations
\begin{equation}\label{approximate_n}
\left\{
\begin{array}{ll}
n(\tilde{T}_j,v_j) & = n(T_j,v_j)+O(r^2) \cr
n(\tilde{T}_j,v_{j+1}) & = n(T_j,v_{j+1})+O(r^2)
\end{array}\right.
\end{equation}
The main purpose of this section is to prove the following result.

\begin{Thm}\label{theorem3} ({\bf Convergence Theorem 2} ) \\
 Let $\Sigma$  be a regular surface and $S=(V,F)$  a triangular surface mesh of  $\Sigma$  with mesh size $r>0$. Consider a smooth vector field $X$  on $\Sigma$, one has, for sufficiently small $r > 0$, and $v \in V$,
 \begin{equation}\label{convergence_divergence}
 \mathrm{Div}_{\Sigma} X(v) = \mathrm{Div}_A X(v) + O(r)
 \end{equation}
 \end{Thm}

 According to the Divergence Theorem on regular surfaces, one has
 \begin{equation}
 \int_W \mathrm{Div}_{\Sigma} X =  \int_{\partial W} < X, \vec{n}>
 \end{equation}
 where the domain $W$  is the union of the geodesic triangles $\tilde{T}_j$ with vertices $v, v_j, v_{j+1}$, $j=0,1,\cdots, n$ and $\vec{n}$ the outer normal vector of $W$. We denote and parametrize the geodesic edge $E_j$  from $v_j$  to $v_{j+1}$  in $\tilde{T}_j$ by $E_j(t)$, $t \in [0,1]$ with $\|E'_j(t)\| = L(E_j)$. Then (\ref{convergence_divergence}) gives
 \begin{equation}
\begin{array}{ll}
\int_W \mathrm{Div}_{\Sigma} X & = \int_{\partial W} <X, \vec{n}> \cr
 & = \sum_{j=0}^{n-1} \int_{E_j} <X, \vec{n}> \cr
 & = \sum_{j=0}^{n-1} L(E_j)\int_0^1 <X(t),\vec{n}(t)>dt
\end{array}
 \end{equation}
where  $L(E_j)$  is the length of the geodesic edge $E_j$. We can
approximate the vectors $X(t)$  and $\vec{n}(t)$  by
\begin{equation}\label{approximate_X_n}
\begin{array}{ll}
X(t) & = (1-t)X(v_j) + tX(v_{j+1}) + O(r^2) \cr
\vec{n}(t) & = (1-t)n(\tilde{T}_j,v_j) + tn(\tilde{T}_j,v_{j+1}) + O(r^2)
\end{array}
\end{equation}
These relations follow from the following easy lemma from Calculus.
\begin{Lem}
Consider a smooth function or vector field $g$  on $[0,a], a> 0$ and a sufficiently small  $r>0$. Then one has, for $t\in[0,1]$,
\begin{equation}
g(tr) = (1-t)g(0)+tg(r)+O(r^2)
\end{equation}
\end{Lem}
Equations (\ref{approximate_n}) and (\ref{approximate_X_n}) imply
\begin{equation}
\vec{n}(t) = (1-t)n(T_j,v_j)  + tn(T_j, v_{j+1}) + O(r^2)
\end{equation}
Note also that the length $L(E_j)$  can be approximated by
\begin{equation}
L(E_j) = \| v_{j+1} - v_j\| (1+O(r^2))
\end{equation}
Hence one obtains
\begin{equation}
\begin{array}{l}
\sum_{j=0}^{n-1}L(E_j)\int_0^1 <X(t), \vec{n}(t)>dt \cr =
\sum_{j=0}^{n-1}\frac{\|v_{j+1} - v_j\|}{6}(1+O(r^2)) \left [ 2<X(v_j),n(T_j,v_j)> \right. \cr
+ 2<X(v_{j+1}), n(T_j,v_{j+1})>  + <X(v_j), n(T_j,v_{j+1})> \cr
\left. + <X(v_{j+1}), n(T_j,v_j)> \right ].
\end{array}
\end{equation}
On the other hand, we also have
\begin{equation}
\begin{array}{ll}
\int_W \mathrm{Div}_{\Sigma} X & = |W| (\mathrm{Div}_{\Sigma} X(v) + O(r) ) \cr
& = \sum_{j=0}^{n-1} |\tilde{T}_j|(\mathrm{Div}_{\Sigma}X(v) + O(r) ) \cr
& = \sum_{j=0}^{n-1} |T_j|(1+O(r^2))(\mathrm{Div}_{\Sigma} X(v) +O(r))
\end{array}
\end{equation}
Therefore we yield
\begin{equation}
\begin{array}{rl}
\mathrm{Div}_{\Sigma} X(v) = & \frac{1}{\sum_{k=0}^{n-1}|T_k|} \left [ \sum_{j=0}^{n-1}\frac{\|v_{j+1} - v_j\|}{6}
( 2<X(v_j), n(T_j,v_j)> \right. \cr
& + 2<X(v_{j+1}, n(T_j,v_{j+1})> + <X(v_j), n(T_j,v_{j+1}> \cr
& \left. + <X(v_{j+1}), n(T_j,v_j)> \right ] + O(r) \cr
= & \mathrm{Div}_A X(v) + O(r)
\end{array}
\end{equation}
and this proves the main theorem (Theorem \ref{theorem3}).

Using the results in subsection 3.2 and this subsection, we can
approximate the Laplace-Beltrami operators on regular surface as
follows. Consider a smooth function $h$  on a regular surface
$\Sigma$ and $S=(V,F)$ a triangular surface mesh of  $\Sigma$ with
mesh size $r>0$. One can use Equations
(\ref{system_gradient})-(\ref{gradient_f}) to define the
approximating gradient $\nabla_A h(v)$ at a vertex $v \in V$. Then
Equation (\ref{divergence_X}) gives the approximating
Laplace-Beltrima operator by
\begin{equation}
\Delta_A h(v) = \mathrm{Div}_A(\nabla_A h)(v)
\end{equation}

Theorems \ref{theorem1} and \ref{theorem3} then give

\begin{Thm}\label{theorem4}( {\bf Convergence Theorem 3} ) \\
 Let $\Sigma$  be a regular surface and $S=(V,F)$  a triangular surface mesh of $\Sigma$  with mesh size $r>0$. Consider a smooth function $h$ on $\Sigma$, one has, for sufficiently small  $r>0$, and $v \in V$,
 \begin{equation}
 \Delta _{\Sigma} h(v) = \Delta_Ah(v) + O(r)
 \end{equation}
\end{Thm}

We can extend $\Delta_A h$ to the whole triangular surface mesh $S$  piecewise linearly in a natural way. Then, the Discrete Conservation Law (Theorem \ref{theorem2}) also holds for the Laplace-Beltrami operators.

\begin{Thm}\label{theorem5} ({\bf Discrete Conservation Law 2 }) \\
Let $S$  be a triangular surface mesh without boundary and $h$ a function on $S$. We have
\begin{equation}
\int_S \Delta_A h = 0
\end{equation}
\end{Thm}

\begin{Rmk}
The error terms $O(r)$, $O(r^2)$ in Theorems \ref{theorem1} -
\ref{theorem5}, can be shown to depend only on curvatures,
injectivity radius \cite{Docarmo2,Jost} of $\Sigma$, vector fields
$X$ and/or functions $h$.
\end{Rmk}

\begin{Rmk}
We also would like to point out that the methods discussed in this
section also work in higher dimensions. Namely, we can also use
these ideas to approximate the gradient, divergence and the
Laplace-Beltrami operators for hypersurfaces in nD Euclidean spaces
with $n \geq 3$. We will discuss these in another paper.
\end{Rmk}

\section{Numerical simulations}

The Laplace-Beltrami operator on a regular surfaces plays an
important role on PDEs. In this section, we shall estimate the
Laplace-Beltrami operators on triangular meshes by our proposed
method and shows some numerical simulations about several important
PDEs on regular surfaces.

\subsection{ Comparisons of Laplacian estimations}
\begin{figure}
{\center
\includegraphics[width=.7\textwidth]{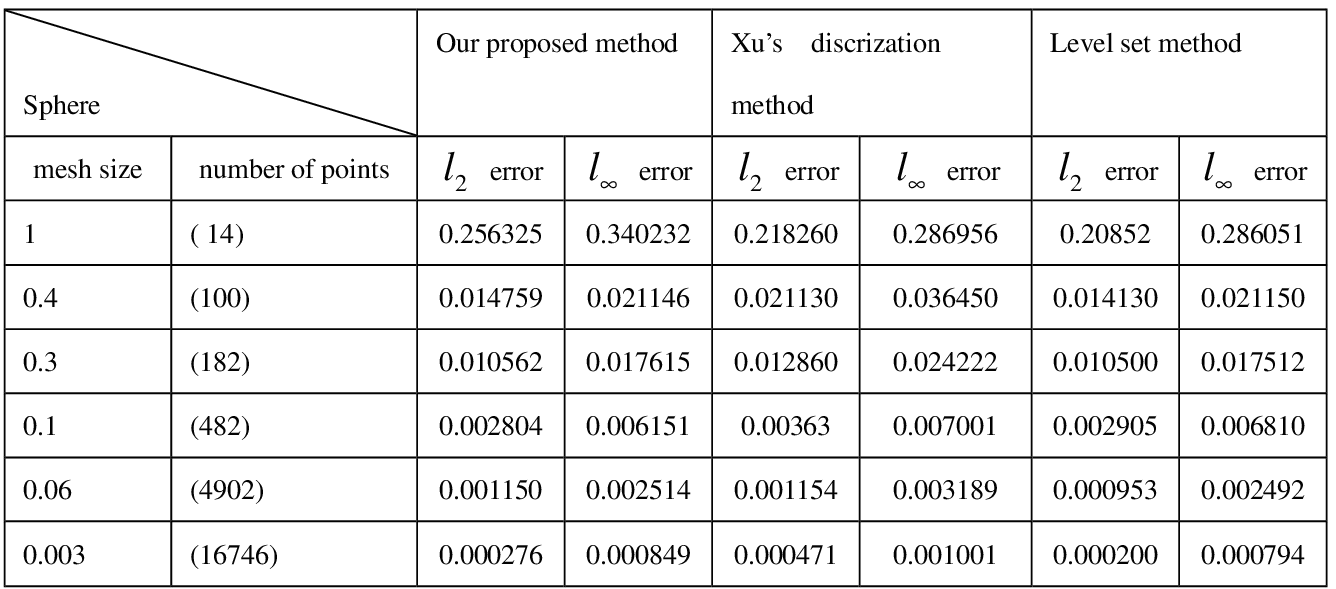}
\caption{ The Laplacian of random polynomial functions on a unit sphere}\label{laplace_sphere}
}
\end{figure}

\begin{figure}
{\center
\includegraphics[width=.7\textwidth]{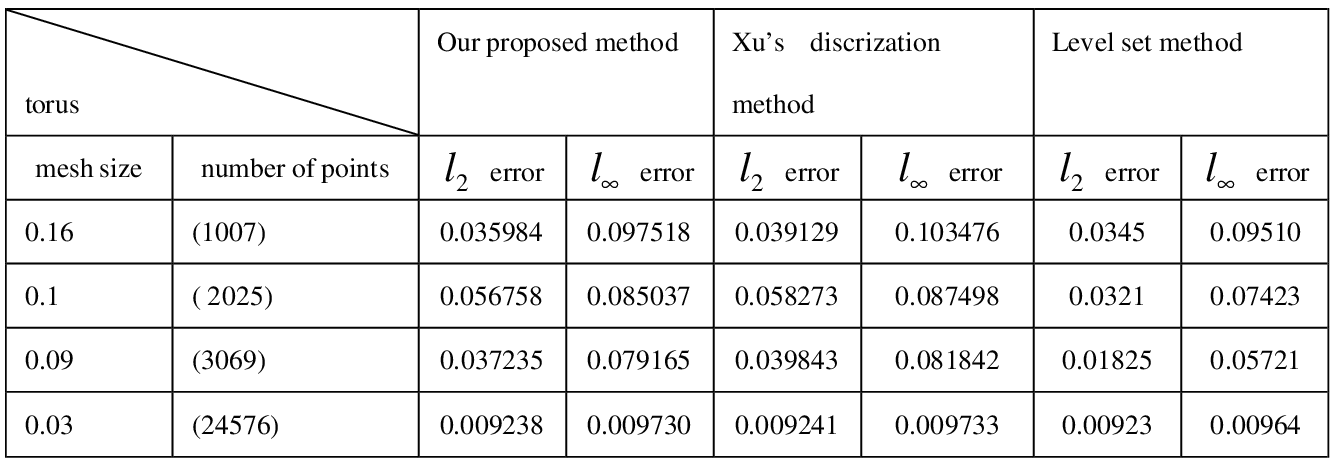}
\caption{ The Laplacian of random polynomial functions on a torus}\label{laplace_torus}
}
\end{figure}

We compare our proposed method, the level set method and some other
discrization methods for estimating the Laplacian of random
polynomial functions of degree less than 5 on a unit sphere and a
torus in figures \ref{laplace_sphere} and \ref{laplace_torus}. Xu's
method is a discretization method proposed in 2004. One can find the
details about Xu's method and Level-set method in \cite{Osher,Xu}.
We choose 10,000 random polynomial functions on these surfaces. The
$l_{\infty}$ and $l_{2}$ errors are used for all vertices on the
triangular mesh. From our simulations, all of these methods are
convergent and comparable.

\subsection{ PDEs on surfaces}
\begin{figure}
{\center
\includegraphics[width=.9\textwidth]{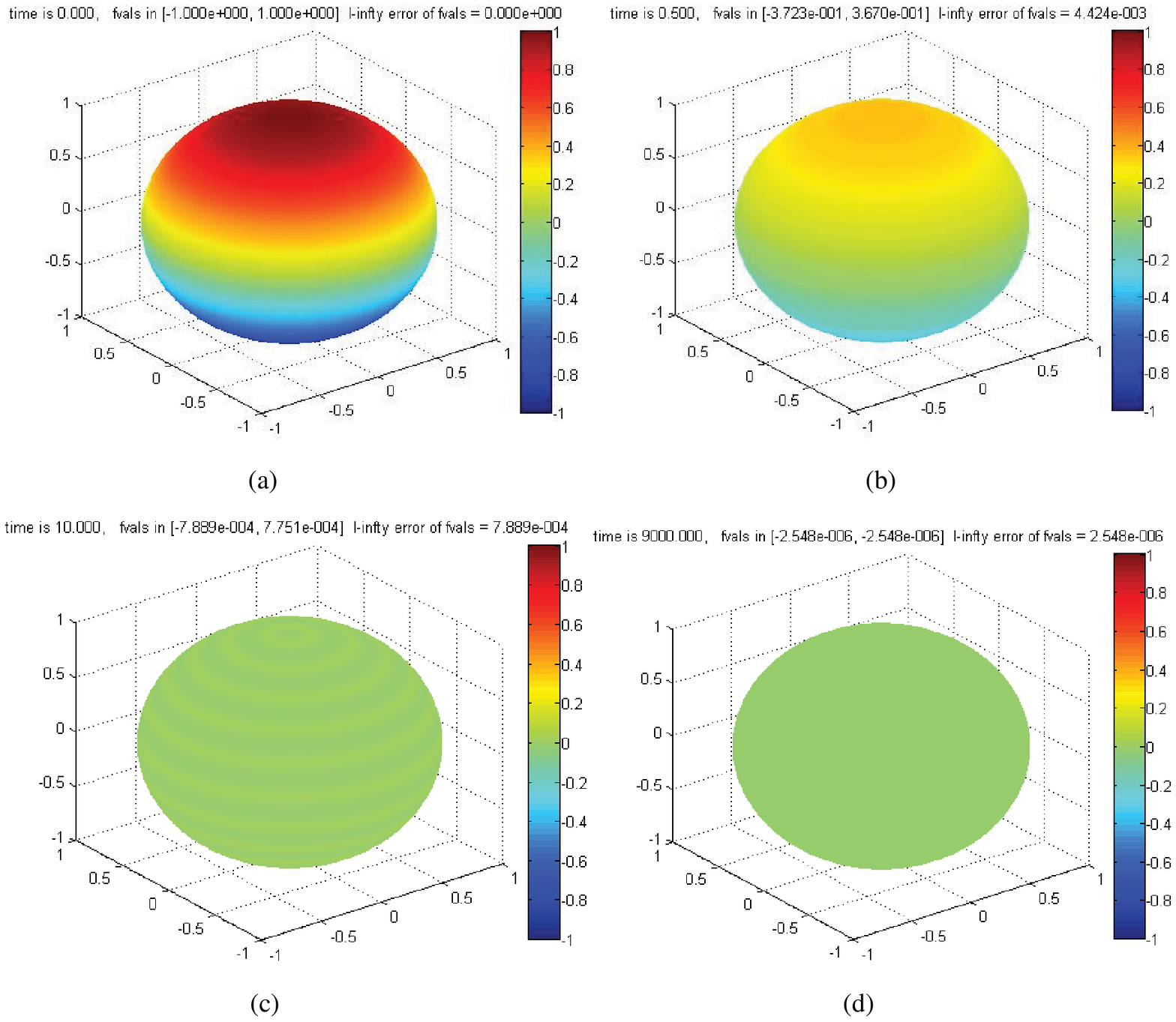}
\caption{ The diffusion equation on a unit sphere}\label{heat_sphere_fig1}
}
\end{figure}

\begin{figure}
{\center
\includegraphics[width=.9\textwidth]{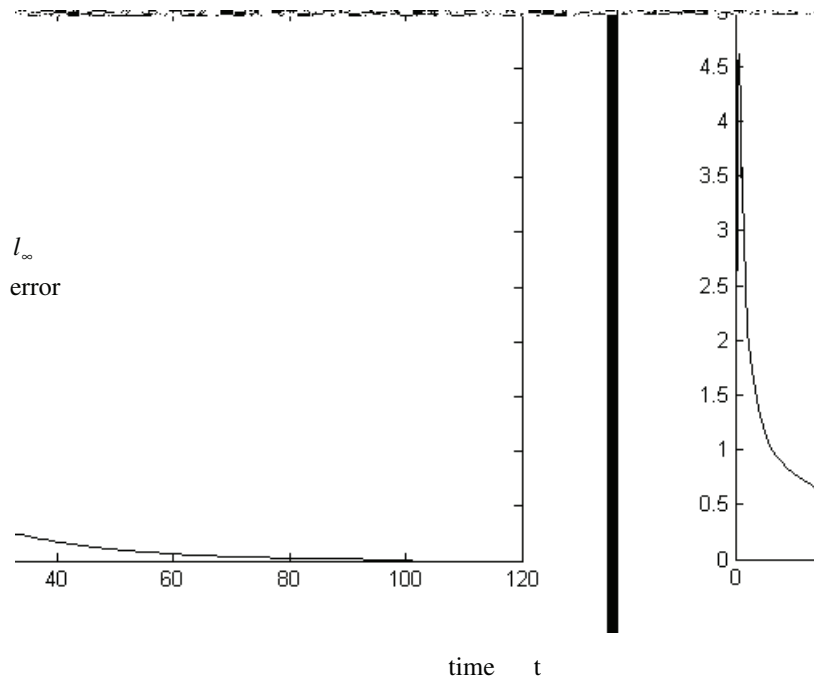}\label{heat_sphere_fig2}
\caption{ The $l_{\infty}$ error of the diffusion equation on a unit sphere}
}
\end{figure}

In this subsection, we show  numerical solutions of some PDEs on
surfaces via our proposed method. First, we consider the diffusion
equation on a sphere
\begin{equation}\label{heat_sphere_equation}
u_t = \Delta_{\Sigma} u
\end{equation}
with the initial condition
\begin{equation}\label{heat_sphere_initial_condition}
u_0(\theta, \eta) = \cos(\eta)
\end{equation}
where $(r, \theta, \eta)$ is the spherical coordinate of the unit sphere.
We calculate
\begin{equation}\label{exact_heat_sphere}
 u(\phi,\theta,t)= \exp(-2t)\cos(\eta)
\end{equation}
as the exact solution of equation (\ref{heat_sphere_equation}) with
initial condition (\ref{heat_sphere_initial_condition}). We compute
the Laplace-Beltrami operator $\Delta_{\Sigma}$ in equation
(\ref{heat_sphere_equation}) by our proposed method and compare our
numerical solution of equation (\ref{heat_sphere_equation}) with the
exact solution (\ref{exact_heat_sphere}). Figure
\ref{heat_sphere_fig1} illustrates the numerical solutions of
equation (\ref{heat_sphere_equation}) at time t=0, 0.5, 10 and
9,000. The "fvals in $[a,b]$" in the figure \ref{heat_sphere_fig1}
means the values $u(t)$ on the surface between $a$ and $b$, and the
"l-infty error" means the $l_{\infty}$ error of our simulations.
Figure \ref{heat_sphere_fig2} gives the $l_{\infty}$ error of our
numerical solutions. Obviously, our numerical solution approaches
the exact solution when the time is large enough. Furthermore, the
integration of $u$, $\int_{\Sigma} u(t;p) dS$, is preserved at all
time.

Next, we solve the fourth order diffusion equation,
\begin{equation}
u_t = - \Delta_{\Sigma}\Delta_{\Sigma} u
\end{equation}
on the sphere with the initial condition
\begin{equation}
u_0 ( r,\theta, \eta) = sin(3\theta)sin(7\eta).
\end{equation}
One can find the details about this equation in Greer's
paper\cite{Greer2}. In our example, the number of triangles on a
triangular mesh is 4096. Figure displays the solution at $t=0$,
$t=0.01$, $t=0.5$ and $t=5$. Obviously, our solution and Greer's
numerical solution\cite{Greer2} are comparable.

\begin{figure}
{\center
\includegraphics[width=.9\textwidth]{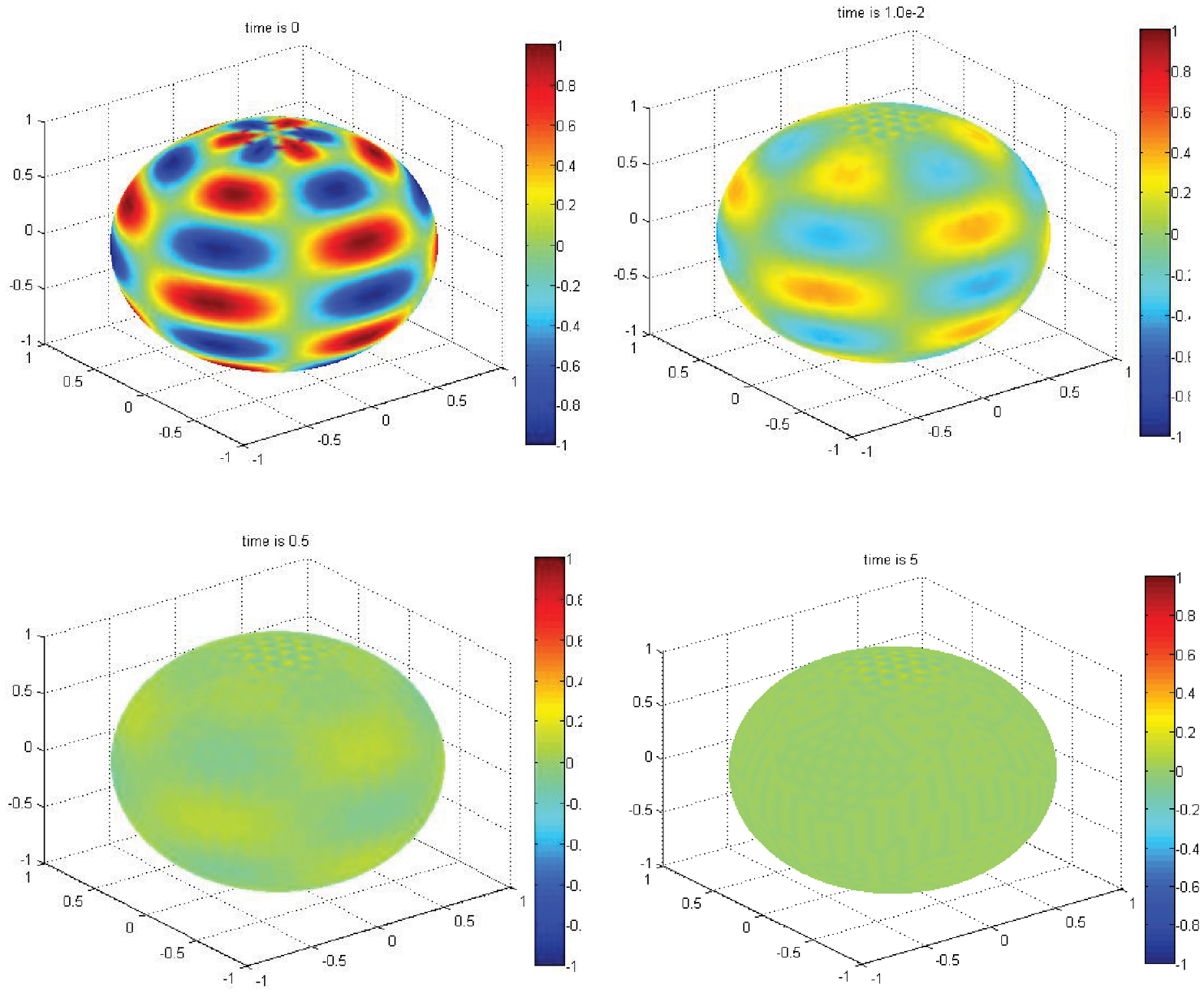}\label{four_sphere_fig}
\caption{ Linear fourth order diffusion on a unit sphere.}
}
\end{figure}

\begin{figure}
{\center
\includegraphics[width=.7\textwidth]{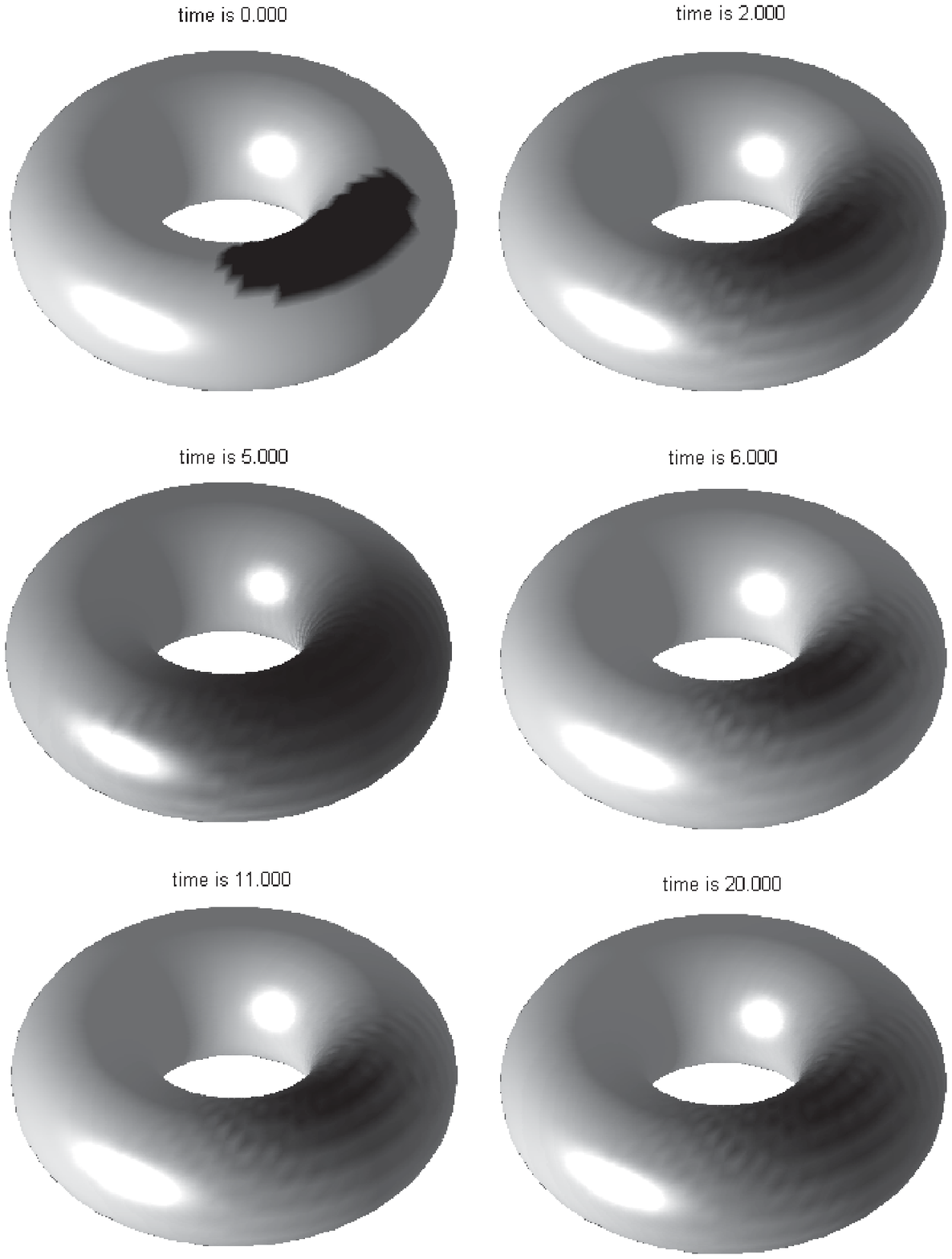}
\caption{ Allen equation on a torus}\label{allen_f1}
}
\end{figure}

For our final example, we compute the Allen-Cahn equation,
\begin{equation}\label{allen}
u_t = \epsilon^2 \Delta_{\Sigma} u + u^3 - u,
\end{equation}
with the initial condition
\begin{equation}\label{allen_initial}
f(\theta, \eta) = \left \{ \begin{array}{ll} 1 & \sqrt{(\theta-\frac{\pi}{2})^2 + (\eta-\frac{\pi}{2})^2 } \leq \frac{4}{5} \cr
-1 & \mbox{ otherwise } \end{array}\right.
\end{equation}
on a torus,
\begin{equation}
\mathbf{x}(\theta, \eta)= \left ( (\frac{1}{2}\cos\eta +1 )\cos \theta, (\frac{1}{2}\cos\eta + 1)\sin\theta, \frac{1}{2}\sin\eta \right ).
\end{equation}
Figure \ref{allen_f1} shows the results. Again, our results and
Greer's numerical solutions \cite{Greer1} are equall, well.

\section{Conclusion}
Our proposed method is a new discretization method for estimating
the divergence of a vector field on surfaces. The convergence ratio
of our proposed method is as good as the other well-known
convergence methods for estimating the Laplace-Beltrami operators.
Almost all of other methods does not obey the divergence theorem,
however  our proposed method does.  That is,  our proposed method
for estimating the Laplace-Beltrami operator on the heat equation
have the conservation property. In the near future, we shall use our
proposed method to improve more partial differential equations, such
as the Navier-Stokes equation, on regular surfaces, triangular
meshes and general manifolds of dimension $n \geq 3$.

\section*{Acknowledgements}
This paper is partially supported by NSC, Taiwan.


\end{document}